# Beyond the NCA: New Results for the Spectral Properties of the Anderson Model

Frithjof B. Anders
Institut f. Festkörperphysik, Techn. Hochschule Darmstadt
Hochschulstr. 6, D-64289 Darmstadt, Fed. Rep. of Germany

11. August, 1994

## Abstract

In the framework of direct perturbation theory a fully self-consistent approximation beyond the well known NCA will be presented for the Anderson Model. The resummation of a class of skeleton diagrams up to infinite order in $V$ includes all contribution up to the order $O(1/N^2)$ ( $N$ = degeneracy of the magnetic state). Qualitative improvements in maintaining local Fermi-Liquid relations and one-particle spectral properties in comparison to the well known NCA will be reported. The location and temperature dependence of the AS-resonance for the case $N = 2$ is found to be rather close to the chemical potential in excellent agreement with Friedel's sum rule; the static magnetic susceptibility exhibits the same $N$-dependence as the exact *Bethe-Ansatz* solution.

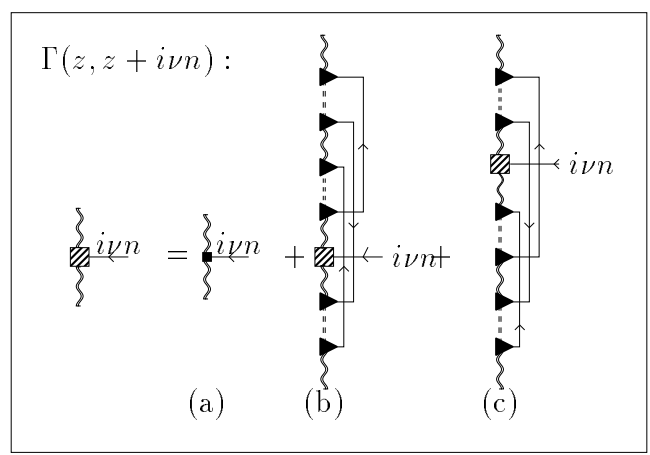

Figure 1: Diagrammatic representation of the self-consistent magnetic vertex function $\Gamma(z, z + i\nu n)$. The wriggle lines represent the singly-occupied state, the dashed lines the unoccupied states and the band electrons are drawn with solid lines.

## 1 Theory

More than ten years ago the first evaluations of a self-consistent infinite order perturbational theory of the Anderson impurity model at infinite $U$ with respect to the hybridization, the NCA, had been reported [2, 3]. It furnishes in particular the correct low-energy scale of the problem, Wilson's Kondo temperature $T_K$. Due to an imbalance of particle and hole excitation processes the NCA contains a pathology at very low energies $T_{path} \ll T_K$, which e.g. shows up as a residual spike in one-particle spectra at the chemical potential $\mu=0$. Resummation of *all* skeleton diagrams yields a set of coupled inte-

gral equations

$$\begin{aligned}\Sigma_0(z) &= \sum_{m=1}^{N} |V|^2 \int_{\infty}^{\infty} de \rho(e) f(e) \\ &\quad P_m(z+e)\Delta_m(z, z+e) \\ \Sigma_m(z) &= |V|^2 \int_{\infty}^{\infty} de \rho(e) f(-e) \\ &\quad P_0(z-e)\Delta_m(z, z-e) \ .\end{aligned} \quad (1)$$

$\rho$ denotes the density-of-states (DOS) of unperturbed conduction electrons. Viewed from an expansion scheme in the inverse of a degeneracy $N$ of the magnetic local state all skeleton diagrams can be classified in orders of $O(1/N)$. $\Delta(x,y)=1$ is the leading approximation of order $O(1/N)$ and recovers the NCA. We have determined the vertex correction self-consistenly up to $O(1/N^2)$[5]. The vertex correction includes particle-particle und hole-hole exitations which are considered as essential to maintain local Fermi-liquid relations. In analogous way, a self-consistent magnetic vertex-function $\Gamma(z, z+i\nu_n)$ has been obtained, Fig.1, to calculate the local magnetic susceptibility $\chi_f(i\nu_n)$

$$\chi_f(i\nu_n) = -\frac{1}{Z_f} \oint_C \frac{dz}{2\pi i} e^{-\beta z} P_m(z) \quad (2)$$
$$\cdot P_m(z+i\nu_n)\Gamma(z, z+i\omega_n) \ .$$

in its natural units $N\mu_B^2 j(j+1)/3$. Fig. 1(a) resembles the NCA contribution and the parts of Fig. 1 (b+c) improve $\chi_f(i\nu_n)$ up to $O(1/N^2)$.

## 2  Results and Discussion

A major improvement in the one-particle spectra by the $O(1/N^2)$-theory (Post-NCA) is presented in Fig.2. For a study of the Kondo regime we choose a band of width $W=20\Delta$ ($\Delta=\pi V^2 \mathcal{N}_F$ being the Anderson width for hybridization $V$ and DOS $\mathcal{N}_F=1/W$) with constant DOS and smooth-edged cut-offs, an unperturbed local level position $\Delta E = -3\Delta$. For the critical degeneracy $N = 2$, where higher order correction in $1/N$ contribute significantly, a large shift of the AS-resonance towards the chemical potential and a reduction of its height can be reported here. This is in accordance with the DOS-rule at $T=0$, which demands $\rho_{fm}(0) = \frac{1}{\Delta\pi}\sin^2(n_f \frac{\pi}{N}) \approx 0.96\frac{1}{\pi\Delta}$ for the present set of parameters ($n_f \approx 0.9$). The validity of the DOS-rule is directly related to the validity of Friedel's sum rule: a cut-down from 20% error in NCA to 3-6% is found in the numerical data. The peak position in the NCA is much more temperature dependent, and the temperature-dependent height of the AS-resonance clearly exceeds the unitary limit value of $\frac{1}{\pi\Delta}$ already at $T \sim T_K$. As is to be expected from the nature of the pertubational approach, no approximation can perfectly redistribute the original sharp spike near $\mu$ generated by the class of most divergent diagrams so that also the PNCA has a remaining deficiency.

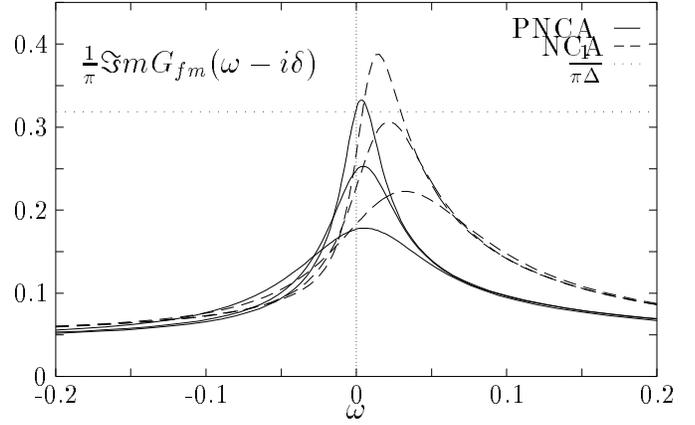

Figure 2: Comparison of the local one-particle spectra at T=0.5, 1.0, 2.0$T_K$ in NCA (dashed) and PNCA (solid lines) in the vicinity of the chemical potential. The lower curves correspond to higher termperatures.

We also succeeded in calculating the static magnetic susceptibility $\chi(T) \equiv \chi(i\nu_n = 0)$ in-cluding higher-order vertex corrections. While $\chi^{NCA}(T)$ shows a rather poor tendency to saturate below $T_K$ for $N = 2$, Fig.3, $\chi^{PNCA}(T)$ calculated with Equ.(2) has overcome this problem, which is related to incorrect threshold exponents of the ionic propagators $P_M(z)$. From the exakt *Bethe Ansatz* results it is known, that $\chi(T)$ exhibits a maximum at $T \sim 0.4 T_K$ for $N > 3$. In Fig.4 excellent agreement with the *Bethe Ansatz* results [4] could be obtained with the parameters $N = 2, \Delta E = -3\Delta$, $N = 4, \Delta E = -5\Delta$ and $N = 6, \Delta E = -7\Delta$. The maximum reproduced in $\chi(0)$ is not accessible to the NCA for $N = 4$. It emphasizes the major improvement provided by the method presented here.

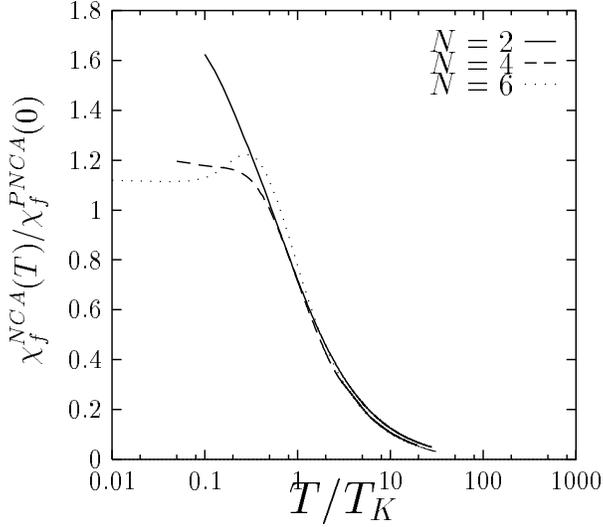 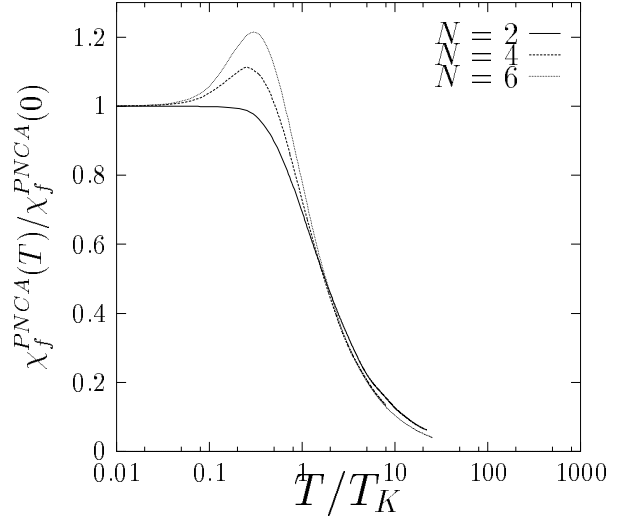

Figure 3: Static magnetic susceptibility for $N = 2, 4, 6$ vs $T/T_K$ in NCA. The parameters are given in the text.

Figure 4: Static magnetic susceptibility for $N = 2, 4, 6$ vs $T/T_K$ in PNCA. Parameters are the same as in Fig.3 and given in the text

The essential progress, however, not only lies in the apparent order of magnitude improvement, compared to the NCA, but also in a stabilisation of the low-energy regime as evidenced by the much smoother temperature dependence. Our method is used to reexamine and improve upon existing lattic theories like the LNCA.

*** 

I would like thank Prof. N. Grewe for many fruitful discussions and encouragement.